\documentclass[%
    twocolumn,
    superscriptaddress,
    aps,
    prxquantum,
    longbibliography
]{revtex4-2}

\usepackage[utf8]{inputenc}
\usepackage{amsmath, amssymb, amsfonts}
\usepackage{braket}
\usepackage{bm}
\usepackage{graphicx}
\usepackage{xcolor}
\usepackage{booktabs}
\usepackage{tabularx}
\usepackage{hyperref}
\usepackage[ruled,vlined]{algorithm2e}
\usepackage{makecell}
\hypersetup{colorlinks=true, linkcolor=blue, citecolor=blue, urlcolor=blue}

\newcommand{\EE}{\mathbb{E}}
\newcommand{\Var}{\mathrm{Var}}

\newcommand{\comp}{\mathcal{C}}
\newcommand{\noise}{\mathcal{N}}

\newcommand{\evec}{\mathbf{e}}
\newcommand{\loss}{\mathcal{L}}
\newcommand{\perm}{\sigma}

\DeclareMathOperator{\err}{err}

\begin{document}

\title{Balanced Routing for Symmetric Quantum Circuits}

\author{Samuel Punch}
\affiliation{University College Cork}

\date{\today}

\begin{abstract}
Quantum programs must be mapped to physical chips with restricted connectivity. Compilers insert SWAP operations to route distant interacting qubits, incurring depth and error penalties. For programs containing cyclic symmetries, routing overhead breaks theoretical symmetry because identical logical roles experience unequal shuffling. This imbalance is commonly attributed to hardware topology alone. We give a sharper, two-level account. Whether a given qubit patch can host a balanced assignment is fixed by its shape; but on any patch that admits balance, the imbalance that actually appears is set by the logical-to-physical assignment rather than the wiring, and a balanced assignment can distribute routing cost perfectly evenly at no extra depth.

Through exhaustive search on a 57-qubit ``heavy-hex'' lattice, we prove these topological constraints exactly. For a four-part ring, 108 of 124 connected patches admit a cost-free balanced assignment, with the 16 exceptions being star-shaped. For a six-part ring, cost-free balance is impossible on compact patches. For a fully connected four-part symmetry, balance is structurally impossible at any depth.

Simulations using realistic error rates matched to measured devices show that, relative to the worst-case concentrated assignment, balanced assignments reduce symmetry-breaking by 
\textbf{92.7\% (95\% CI [+89.8\%, +95.3\%])} for the raw metric and 
\textbf{87.0\% (95\% CI [+79.6\%, +94.1\%])} for the decoherence-corrected measure ($p = 2.45 \times 10^{-32}$). 
Substrate error heterogeneity accounts for at most \textbf{10.8\%} of this effect. Notably, switching to the compiler's highest generic optimization level did not produce a statistically significant change in routing imbalance, highlighting the need for targeted symmetry-aware passes.

We conclude that when patch geometry permits, routing imbalance is a compiler choice rather than a hardware limitation. Consequently, symmetry-aware assignment should be a primary objective for both compiler optimization and quantum chip design.
\end{abstract}
\maketitle

\section{Introduction}
A quantum compiler turns an abstract program into instructions a specific chip can run. Its main job is placement and routing: deciding which physical qubit holds which piece of the program, and inserting extra operations wherever two pieces that must interact are not wired together. Every extra operation costs depth (roughly, running time) and adds error, so good compilation matters.

This paper is about programs with a symmetry that hardware quietly destroys. Some circuits are built so that a group of their parts are interchangeable, rotating or relabelling those parts should not change the answer. The clearest example is an \emph{equivariant} quantum circuit, a circuit deliberately constrained so that certain rearrangements of its inputs leave the output unchanged, much as a good weighing scale reads the same value no matter how you rotate the object on the pan~\cite{nguyen2024, schatzki2024}. These circuits are attractive because the symmetry constraint improves how well they train and generalise~\cite{schatzki2024, west2024}. The catch is that those guarantees assume perfect, noise-free execution. Once the circuit is compiled onto real hardware, the guarantee and the actual behaviour can diverge, and until now that gap was not well understood.

Prior work by T\"{u}ys\"{u}z \emph{et al.}~\cite{tuysuz2024} showed that hardware noise behaves like a force that breaks the symmetry. That analysis treated the noise as the same everywhere on the chip. It noted, but did not pursue, a separate compilation-level worry: the operations needed to protect the symmetry might not survive compilation. We take up exactly that worry and pin down the mechanism.

Our starting point is a fact that sounds backwards. Compilation does not break the symmetry by changing what the circuit computes, a circuit routed with exact SWAPs still computes the symmetric answer. The symmetry breaks because the chip is physically uneven: interchangeable parts of the ring get mapped to different physical qubits along different shuffling paths, so they accumulate different amounts of error. Our central result is that, for a fixed patch of qubits, this imbalance is set by the assignment of parts to qubits, not by the wiring itself, while whether any balanced assignment exists is set by the patch's shape.

We establish this by exhaustive search on the heavy-hex chip layout used by current superconducting hardware. Our concrete findings, each a proof over all possibilities rather than a sample, are:
\begin{itemize}
    \item \textbf{Four-part ring ($C_4$):} $108$ of $124$ connected qubit patches allow a balanced assignment at no extra depth. The $16$ that fail are all star-shaped patches (one central qubit wired to three others).
    \item \textbf{Six-part ring ($C_6$):} no compact patch allows balance for free; balance exists only at extra depth.
    \item \textbf{Fully connected four-part symmetry ($S_4$):} no compact patch allows balance at any depth.
\end{itemize}

The result reaches beyond equivariant circuits. To understand the scope of this problem, it is important to recognise where these cyclic ``ring'' structures appear in practice. They are not universal; algorithms requiring all-to-all global entanglement (like the Quantum Fourier Transform) or star-dependent functional oracles (such as Grover's algorithm) generally lack them. However, rings are a fundamental building block in algorithms that map structured physical systems or regular data. Trotterized physics simulations (such as Ising or Heisenberg models on a 2D grid), grid-based QAOA for combinatorial optimization, and circular hardware-efficient ansatzes all rely heavily on periodic boundary conditions and closed local loops. Because these algorithmic families rely on repeated cyclic blocks, we expect the same assignment dependence wherever such a block is compiled onto restricted connectivity, though we prove it here only for isolated four- and six-part rings, not for full multi-block circuits. Three practical consequences follow.
\begin{enumerate}
    \item \textbf{Compilers} should spread routing cost evenly across a cyclic block, rather than leaving it to a generic optimiser.
    \item \textbf{Chip designers} should treat ``can this layout host balanced assignments'' as a design metric.
    \item \textbf{Benchmarks} for cyclic workloads should report routing imbalance, not just average gate count.
\end{enumerate}

\section{Related Work}
We place our contribution against three lines of prior work; full detail is in the references.

\paragraph*{Noise that breaks symmetry.} Reference~\cite{tuysuz2024} models chip noise as uniform across all qubits. Because it does not track \emph{where} on the chip each ring part runs, it cannot see effects that come from uneven error rates or from unequal routing cost. Those are exactly the effects we isolate.

\paragraph*{Layout that ignores roles.} Circuit builders such as TopGen~\cite{topgen} and topology-aware synthesis~\cite{topas} minimise total depth or two-qubit gate count. They do not care which physical qubit carries which logical role. As we show, cheapest-to-route is not the same as most-symmetric.

\paragraph*{Picking low-error qubits.} A family of methods~\cite{mapomatic,murali2019,tannu2019} chooses physical qubits by their measured error rates. This helps average error but cannot fix cyclic imbalance: the fix is how you assign roles to a patch, not which patch you pick.

\section{ORIGIN OF ROUTING IMBALANCE}
\label{sec:mechanism}
This section explains, in one mechanism, why a symmetric circuit becomes asymmetric on hardware. The short version: rotating the circuit's input is equivalent to relabelling which qubit plays which role, so unequal routing across those roles shows up as a broken symmetry.

Consider a cyclic dependency of order $n$: a ring of $n$ logical qubits, each meant to play the same role. If the input is transformed by a symmetry operation $x\mapsto R(g)x$, the ideal answer does not change:
\begin{equation}
\loss(R(g)x)=\loss(x).
\label{eq:invariance}
\end{equation}
So any spread we measure across the ring is direct evidence of a hardware-induced deviation.

\paragraph*{The duality lemma (why rotating the input equals relabelling the chip).} For the symmetries we study, the symmetry operation is just a permutation of qubits, $U_g=P_{\perm(g)}$. Substituting it into a fixed hardware placement $\pi$ gives an exact identity:
\begin{equation}
\underbrace{\loss(R(g)x)\ \text{on placement } \pi}_{\text{what we measure}}
\;=\;
\underbrace{\loss(x)\ \text{on placement } \pi\!\circ\!\perm(g)}_{\text{what we score}} .
\label{eq:duality}
\end{equation}
In words: measuring the answer as we rotate the input is the same as leaving the input fixed and relabelling which physical qubit holds which role. This is why the measured spread is really a statement about assignment.

\paragraph*{Adding up the error along each role.} Write the real (noisy) run as ideal compilation followed by noise, $\comp_\pi=\noise\circ\comp_\pi^{\,\mathrm{ideal}}$. The symmetry survives only if the noise treats every ring role the same. To measure how it treats each role $g$, we add up the error of every gate that role passes through in the compiled circuit:
\begin{equation}
[\evec_\pi]_g=\sum_{\text{gates } h\in \comp_{\pi\circ\perm(g)}}\err(h).
\label{eq:evec}
\end{equation}
Adding over gate \emph{instances}, not just the set of qubits touched, is essential. A count that ignores how many times each qubit is used would report zero spread precisely in the case we care about, where roles share the same qubits but differ in gate count.

\section{QUANTIFYING ROUTING ASYMMETRY}
\label{sec:metric}
We need a single number that is zero when the ring is perfectly balanced and grows as it becomes uneven. We use the spread (variance) of the per-role measurement across the ring.

Let $\hat\mu_g$ be the measured value of a $\pm1$ observable for ring role $g$, estimated from $S$ measurement shots. The raw spread is
\begin{equation}
s^2=\frac{1}{n}\sum_{g\in G}\big(\hat\mu_g-\bar\mu\big)^2,
\qquad \bar\mu=\frac{1}{n}\sum_{g\in G}\hat\mu_g.
\label{eq:orbitvar}
\end{equation}
Because each $\hat\mu_g$ comes from a finite number of shots, this raw spread is slightly too large, finite sampling adds apparent variance even when the true variance is zero. Writing $\hat\mu_g=\mu_g+\epsilon_g$ with $\Var(\epsilon_g)=(1-\mu_g^2)/S$, the inflation is $\tfrac{n-1}{n}\bar v$. We subtract its unbiased plug-in estimate, which replaces the unknown $\mu_g$ by the measured $\hat\mu_g$ and $S$ by $S-1$:
\begin{equation}
\boxed{\;\hat{E}=\EE_x\!\left[\,s^2-\frac{n-1}{n}\,
\overline{\left(\frac{1-\hat{\mu}_g^2}{S-1}\right)}\,\right]\;}
\label{eq:metric}
\end{equation}
where the bar averages over $g$. This corrected number is, on average, exactly the true spread. We report it without clipping.

As a circuit loses signal to decoherence, all outcomes shrink toward zero, which can shrink the spread for the wrong reason. To separate genuine balance from fading signal, we also report a decoherence-corrected version,
\begin{equation}
E_{rel}=\frac{\hat E}{\bar{\mu}^2},
\label{eq:erel}
\end{equation}
and only trust it when the signal is above the noise floor ($|\bar{\mu}|>3\sigma_{\bar{\mu}}$).

\section{SEARCH AND TOPOLOGICAL CONSTRAINTS}
\label{sec:diagnostic}
This section shows, first by a worked example and then by checking every case, that the imbalance is an assignment choice on a given patch---and maps out exactly which patches allow a balanced choice.

\paragraph*{The worked example.} A four-part ring needs the four links $(0,1),(1,2),(2,3),(3,0)$. Put the four qubits on a straight path $p_0\!-\!p_1\!-\!p_2\!-\!p_3$ and assign them in order. Three links are between neighbours, but the fourth has to reach from one end of the path to the other, so all the shuffling piles onto one ring element. Compiled with Qiskit 2.5.1, the four rotated versions cost $[10,10,16,16]$ two-qubit gates, a spread of $\Var(cz)=9.0$. (Older compiler versions reported $12.0$ for this same worst case; the exact number depends on the routing algorithm, but the effect does not.) This is the most uneven assignment possible.

\paragraph*{Checking every assignment.} There are $24$ ways to assign four roles to four qubits on a path; after removing rotations and reversals of the ring (which are genuinely the same), six distinct classes remain. We tag each by its ring-distance profile, the list of gaps between consecutive roles:
\begin{multline}
\text{profile}=(\text{dist}(\pi(0),\pi(1)),
\text{dist}(\pi(1),\pi(2)),\\
\text{dist}(\pi(2),\pi(3)),
\text{dist}(\pi(3),\pi(0))).
\end{multline}
On the 57-qubit chip, the $108$ path-shaped patches all give the pattern in Table~\ref{tab:profiles}.

\begin{table}[htbp]
\centering
\small
\caption{On a straight path, the assignment, not the hardware, sets the imbalance. The balanced and concentrated assignments use the same qubits and the same average depth; only the spread differs. Values from Qiskit 2.5.1; older versions reported 12.0 for the concentrated case.}
\label{tab:profiles}
\begin{tabular}{@{}lccc@{}}
\toprule
\textbf{Ring profile} & \textbf{Gap sum} & \makecell{\textbf{Mean two-}\\ \textbf{qubit gates}} & \textbf{Spread} \\
\midrule
(1,2,1,2) balanced & 6 & 13.00 & 0.00 \\
(1,1,1,3) concentrated & 6 & 13.00 & 9.00 \\
(1,2,3,2) mixed & 8 & 14.50 & 2.25 \\
\bottomrule
\end{tabular}
\end{table}

\begin{figure}[htbp]
    \centering
    \includegraphics[width=\columnwidth]{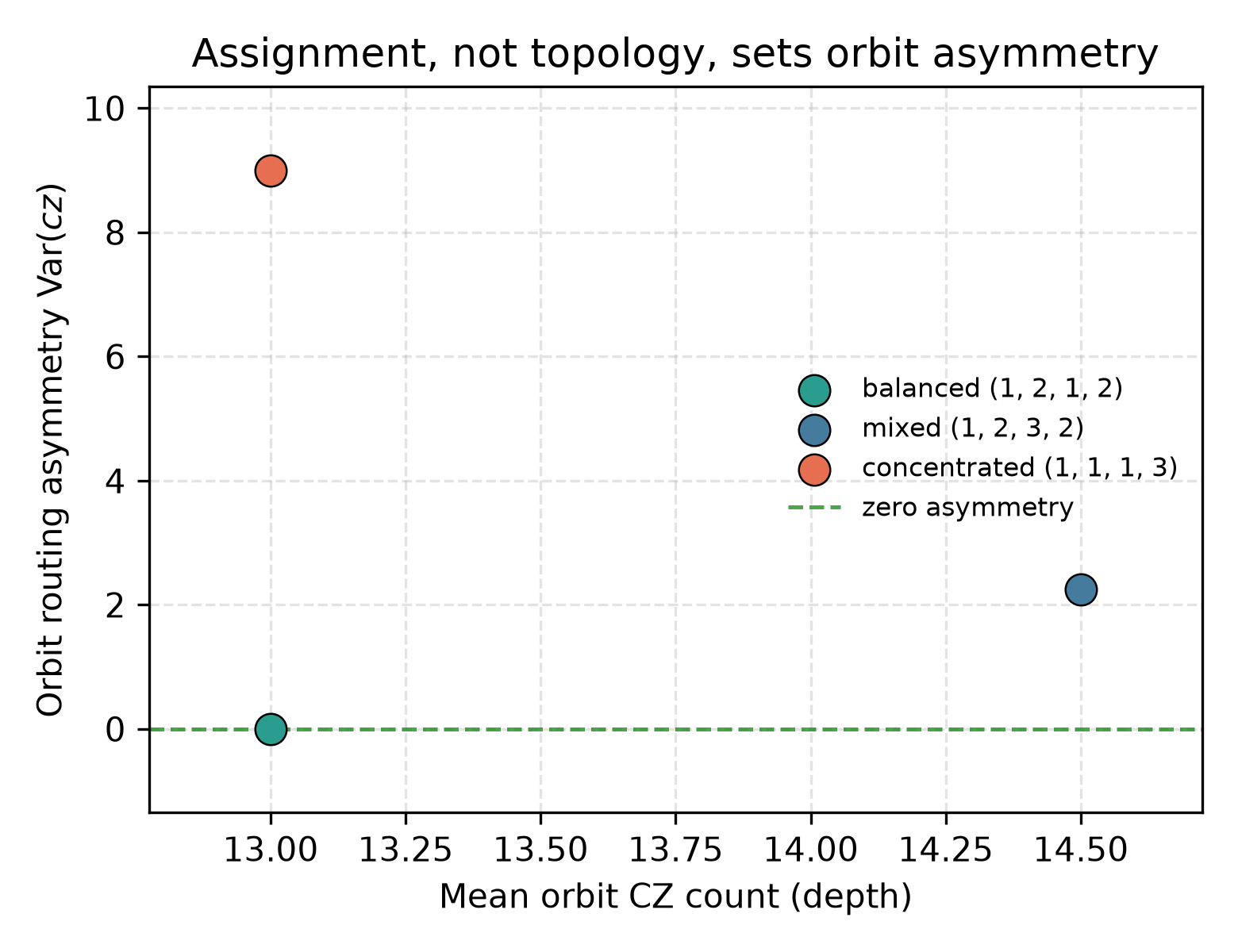}
    \caption{The assignment sets the imbalance at fixed depth. Each point is one way of assigning a four-part ring to the same straight path. The balanced assignment $(1,2,1,2)$ has zero spread; the concentrated assignment $(1,1,1,3)$ has the worst spread ($9.0$ on Qiskit 2.5.1; $12$ on older versions). Both sit at the same average depth ($13$ two-qubit gates) and identical transpiled circuit depth---only the spread differs. In the axis labels, ``orbit'' denotes a ring role, so ``orbit routing asymmetry'' is the per-role spread reported as Spread in Table~\ref{tab:profiles}.}
    \label{fig:assignment}
\end{figure}

The balanced and concentrated rows share the same gap sum and the same average depth. Only the spread differs: one is zero, the other is the worst case. So on any path, a balanced assignment exists and costs nothing extra. 
Crucially, we also confirm that the transpiled circuit depth (number of layers, as reported by \texttt{qc.depth()}) is identical for both assignments; the benefit is thus purely in the distribution of routing cost, not in overall circuit length. Figure~\ref{fig:assignment} shows the three classes.

\paragraph*{Proving it for every patch, not just paths.} We next check all \emph{connected} patches of a given size, not only straight paths, using a search that lists every connected patch exactly once. We do not shortcut this by grouping patches that ``look the same,'' because routing cost depends on the exact distances through the chip and is not captured by shape alone; we use a shape fingerprint (a Weisfeiler--Lehman graph hash, a quick label that is usually different for different shapes) only to report how many patches of each shape there are, never to skip any patch in the proof. For each patch we generate every assignment up to the ring's own rotations and reversals. We report two verdicts:
\begin{itemize}
    \item \textbf{Exists at any depth:} is there \emph{any} assignment, on any patch, with zero spread?
    \item \textbf{Free lunch:} does the patch's own cheapest assignment already have zero spread?
\end{itemize}
Table~\ref{tab:exhaustive} gives the verdicts. Because the search is exhaustive, a ``no'' is a proof of impossibility, not a failure to find.

\begin{table}[htbp]
\centering
\small
\caption{Verdicts from exhaustive search over all connected patches. ``Exists'' asks whether balance is reachable at any depth; ``Free lunch'' asks whether it is reachable at the patch's own cheapest depth. Each verdict is complete for the stated chip size.}
\label{tab:exhaustive}
\begin{tabular}{@{}lcccc@{}}
\toprule
\textbf{Symmetry} & \textbf{Patches} & \textbf{Shapes} & \textbf{Exists} & \textbf{Free lunch} \\
\midrule
$C_4$ & $124$ ($d{=}5$) & $2$ & yes & yes ($108/124$) \\
$S_4$ & $124$ ($d{=}5$) & $2$ & no & no \\
$C_6$ & $62$ ($d{=}3$) & $3$ & \makecell{yes \\ ($d{=}5$)} & no \\
\bottomrule
\end{tabular}
\end{table}

\begin{figure}[htbp]
    \centering
    \includegraphics[width=\columnwidth]{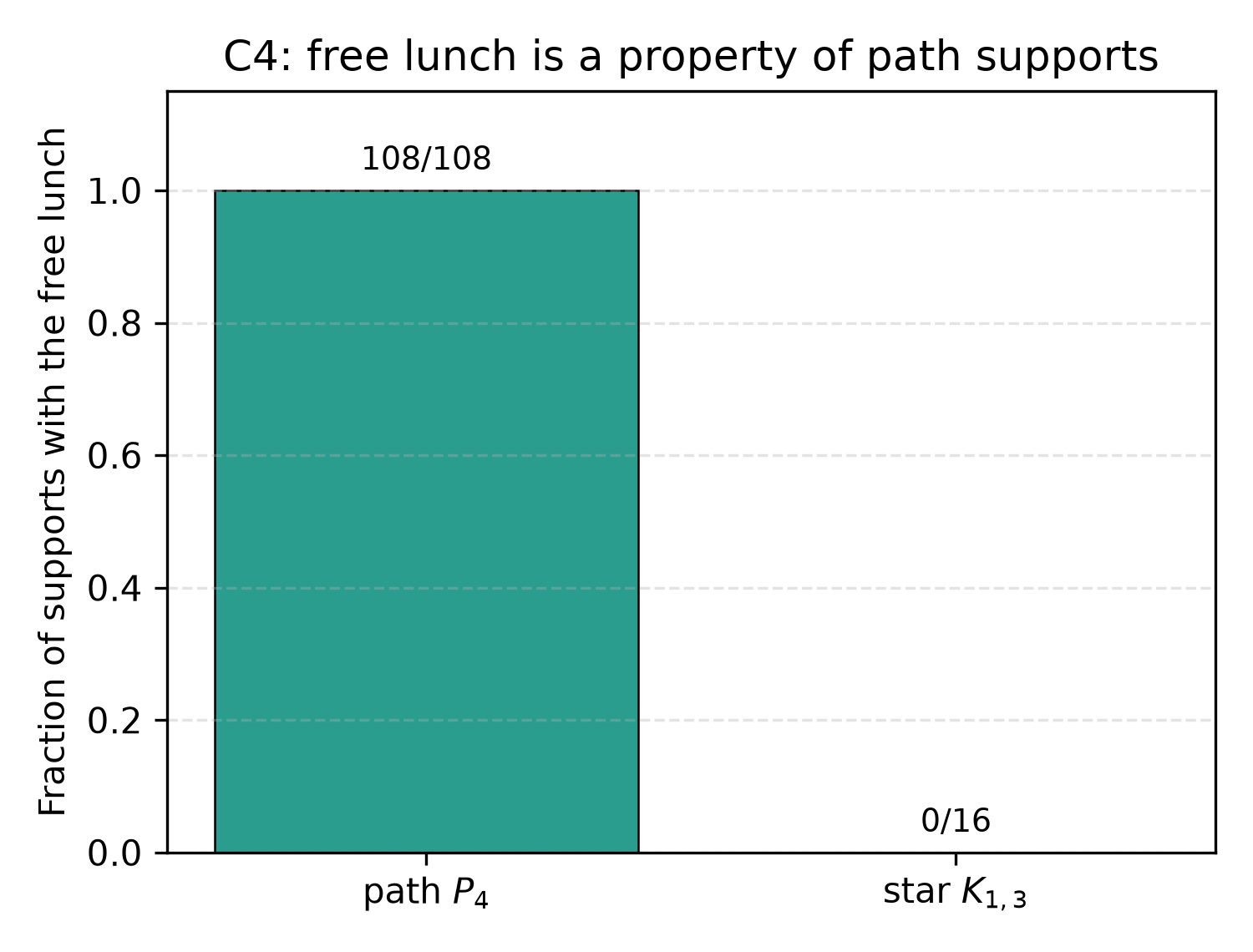}
    \caption{Why $16$ patches fail. Connected four-qubit patches on heavy-hex come in only two shapes: straight paths and stars. All $108$ path patches admit a balanced, cost-free assignment; none of the $16$ star patches do. The free lunch is a property of path-shaped patches. (In the figure, ``support'' is used for a patch.)}
    \label{fig:shapes}
\end{figure}
Three things stand out. First, for the four-part ring the free lunch works on $108$ of $124$ patches; the $16$ failures are exactly the star-shaped patches, where one qubit sits at the centre and cannot be given an even share of the ring (Fig.~\ref{fig:shapes}). 
A star-shaped patch has a central qubit of degree 3, while a four-cycle ring requires each of its four roles to have exactly two neighbours in the distance profile. The central qubit therefore inevitably oversubscribes one role, making the balanced profile $(1,2,1,2)$ unattainable. 
Second, for the six-part ring the free lunch is gone: no compact patch balances at its own cheapest depth, though balance still exists if one pays extra depth (about $27$ two-qubit gates, versus $13$ for the four-part case). Third, for the fully connected four-part symmetry, no patch balances at any depth, because that symmetry demands every part interact with every other, and this sparse chip cannot place four qubits all at equal distance.

\paragraph*{Plain statement of scope.} 
We prove the six-part ``no free lunch'' result on the smaller $19$-qubit chip ($d{=}3$), where the search is fast; the four-part and fully connected results are on the full $57$-qubit chip ($d{=}5$). The exhaustive proof on the $19$-qubit chip already establishes the topological obstruction for compact patches. Extending the verification to the full $57$-qubit layout is a straightforward but longer computation (estimated at half an hour); we explicitly leave this as a scoping limitation of the present exhaustive proof, rather than a gap in the mechanism.

\paragraph*{Routing intervention with generic optimisation.} 
Switching the compiler from its default routing pass (\texttt{optimization\_level=1}) to the most aggressive setting (\texttt{optimization\_level=3}) on general four-qubit placements produced an estimated change in routing imbalance of $-13.5\%$ with a 95\% confidence interval spanning $[-32.6\%, +14.0\%]$. 
Because the interval crosses zero, this result is statistically inconclusive: we cannot claim that the aggressive optimiser systematically worsens the imbalance, nor that it reliably improves it. This underscores that generic depth-minimising passes are not a substitute for targeted symmetry-aware assignment, as their effect on routing imbalance is unpredictable.

\section{Simulation with Realistic Noise}
This section confirms that the balanced assignment actually helps under realistic error, not just in the gate count.

\begin{figure}[htbp]
    \centering
    \includegraphics[width=\columnwidth]{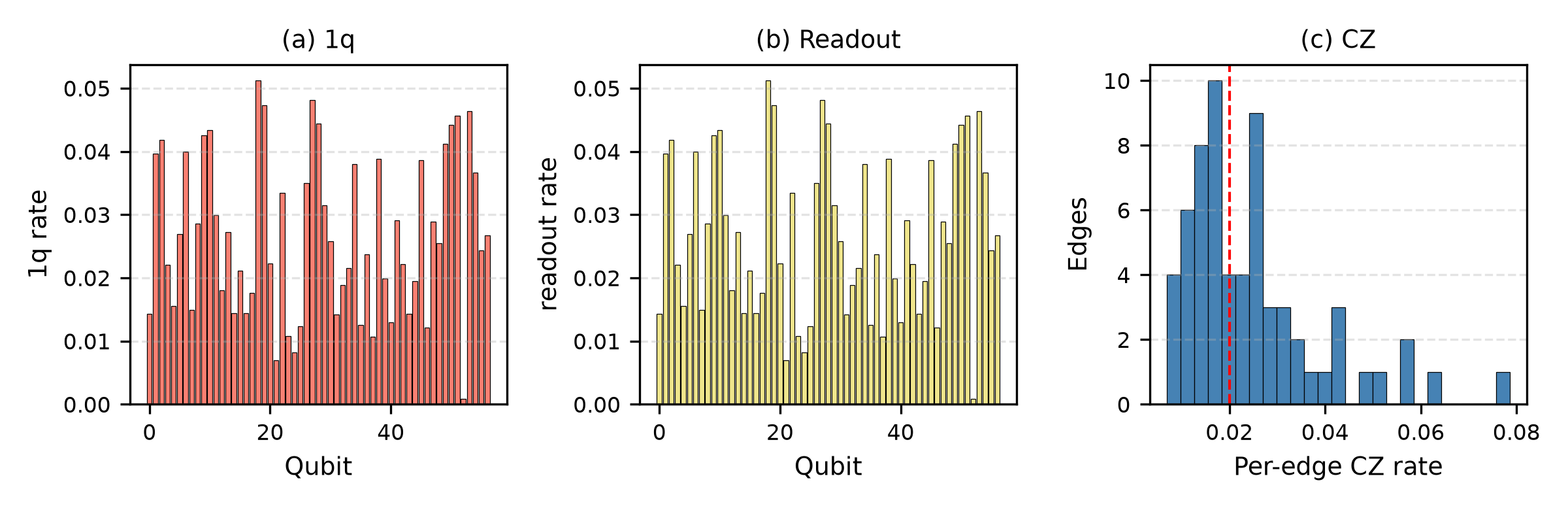}
    \caption{The simulated error field (IBM Heron): single-qubit error rates, readout error rates, and the spread of two-qubit gate error across the chip's links.}
    \label{fig:substrate}
\end{figure}

Setup. We simulate on the 57-qubit chip using error rates drawn to match measured IBM Heron and Eagle devices, with $16{,}000$ measurement shots per setting. Figure~\ref{fig:substrate} illustrates the simulated error distributions across the 57 qubits for single-qubit, readout, and two-qubit gates. We test all 108 path patches, with 24 input probes each and three independent random seeds.

The fair comparison. For each path we compare two assignments that differ in one thing only:
\begin{itemize}
    \item \textbf{Balanced:} zero spread, profile $(1, 2, 1, 2)$.
    \item \textbf{Concentrated:} worst-case spread, profile $(1, 1, 1, 3)$.
\end{itemize}
Both use the same physical qubits, the same average gate count, and the same error exposure. Any difference in the result is due to the spread alone.

\begin{figure}[htbp]
    \centering
    \includegraphics[width=\columnwidth]{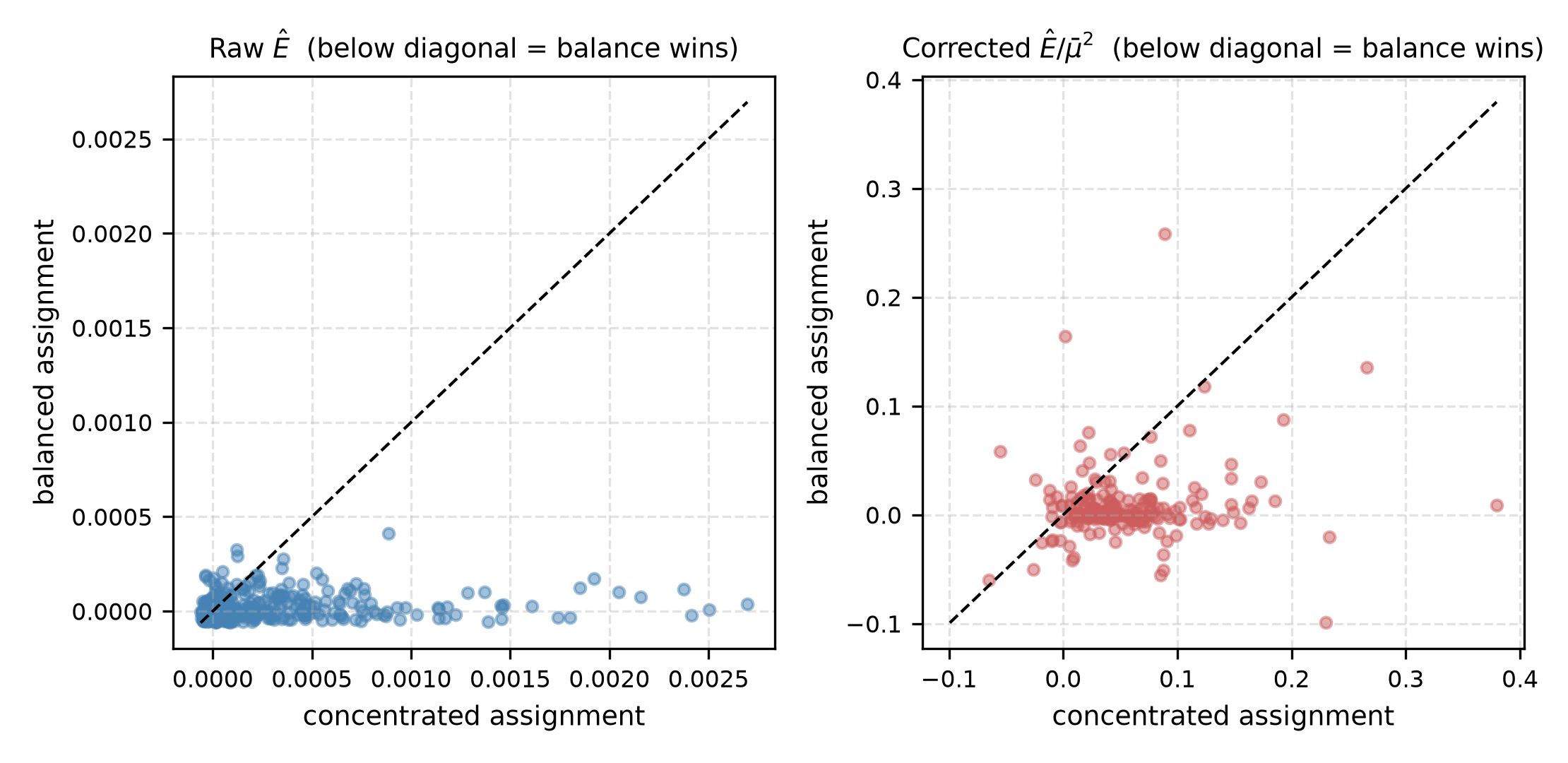}
    \caption{Balanced versus concentrated on the same path, under realistic noise. Each point is one path; points below the diagonal mean the balanced assignment did better. Left: raw symmetry-breaking. Right: the decoherence-corrected version. The balanced assignment wins on the large majority of paths.}
    \label{fig:within}
\end{figure}

Result. The balanced assignment consistently and significantly reduces symmetry-breaking. Figure~\ref{fig:within} plots this comparison across all paths, showing that the majority of points fall below the diagonal. Aggregating over all paths and seeds, the raw metric ($\hat E$) reduction is $92.7\%$ with a $95\%$ confidence interval of $[+89.8\%, +95.3\%]$. The decoherence-corrected metric ($E_{rel}$) shows a reduction of $87.0\%$ with a $95\%$ CI of $[+79.6\%, +94.1\%]$ ($n=212$ paths passing the resolution guard). A paired Wilcoxon test confirms the effect is highly significant ($p=2.45\times10^{-32}$), with the balanced assignment outperforming the concentrated one on $78\%$ of paths.
Because the concentrated profile is the worst possible assignment on a path, these reductions measure the maximum symmetry-breaking that balancing can remove, not the gain over a default compiler: where a generic optimiser lands between the balanced and concentrated extremes is workload-dependent and, as the inconclusive routing-intervention test of Sec.~\ref{sec:diagnostic} indicates, not reliably predictable.

Crucially, a channel-resolved ablation shows that heterogeneity in readout, single-qubit, and two-qubit error rates accounts for at most $10.8\%$ of the matched-mean baseline effect ($95\%$ confidence). The remaining effect is therefore genuinely due to the routing imbalance itself, rather than to uneven qubit quality. Table~\ref{tab:results} summarises the aggregate statistics.

\begin{table}[htbp]
\centering
\caption{Aggregate reduction in symmetry-breaking when using the balanced assignment versus the concentrated assignment on the same physical path, across all 108 paths and 3 seeds ($S=16{,}000$).}
\label{tab:results}
\resizebox{\columnwidth}{!}{%
\begin{tabular}{@{}lccc@{}}
\toprule
\textbf{Metric} & \textbf{Reduction} & \textbf{95\% CI} & \textbf{$p$ (Wilcoxon)} \\
\midrule
$\hat E$ (raw) & 92.7\% & [+89.8\%, +95.3\%] & $2.45\times10^{-32}$ \\
$E_{rel}$ (decoherence-corrected) & 87.0\% & [+79.6\%, +94.1\%] & $2.45\times10^{-32}$ \\
\bottomrule
\end{tabular}%
}
\end{table}

\section{CLOSED-FORM BALANCING RULE}
\label{sec:algorithm}
For the four-part ring, the exhaustive search reveals a rule so simple it needs no search at run time. On a path $p_0,p_1,p_2,p_3$, assign the ring as:
\begin{equation}
\pi(0)=p_0,\quad \pi(1)=p_1,\quad \pi(2)=p_3,\quad \pi(3)=p_2 .
\label{eq:canonical}
\end{equation}
In plain terms: place the first two roles in order, then swap the last two. This produces the balanced profile $(1,2,1,2)$ with zero spread, at the cheapest possible depth ($13$ two-qubit gates). Across all $108$ paths this is the only pattern (up to the ring's rotations and reversals) that reaches zero spread for free. Reversals count as the same because an undirected ring reads the same forwards and backwards.

\begin{algorithm}[H]
\SetAlgoLined
\DontPrintSemicolon
\SetKwProg{Fn}{Function}{}{}
\SetKw{Data}{Input:}
\SetKw{Result}{Output:}
\Fn{\textsc{BalancedFourRing}$(P)$}{
    \Data physical path $P = [p_0, p_1, p_2, p_3]$\;
    \Result assignment $\pi:\{0,1,2,3\}\to P$ and spread $\Var(cz)$\;
    $\pi(0) \gets p_0$ \;
    $\pi(1) \gets p_1$ \;
    $\pi(2) \gets p_3$ \;
    $\pi(3) \gets p_2$ \;
    $\text{profile} \gets [\text{dist}(\pi(i), \pi((i+1) \bmod 4)) \ \text{for } i = 0,1,2,3]$\;
    $\Var(cz) \gets \text{Variance}(\text{profile})$\;
    \Return{$\pi$, $\Var(cz)$}
}
\end{algorithm}

The rule runs in constant time and adds no depth. It only reorders the assignment.

\paragraph*{Why we stop at four.} We do not claim a simple rule for larger rings, and we say so plainly. For the six-part ring there is no free balanced assignment on a compact patch, so any rule would have to trade depth for balance. Finding the best such trade is a harder placement problem that we leave open. In the meantime, a compiler can call the exhaustive search directly for small rings, or accept a depth cost for larger ones.

\section{Discussion}
Our main message is a two-level picture. On a patch that admits balance, the imbalance is a choice the compiler makes, and it can be removed for free. But whether any patch admits balance is decided by its shape: for the four-part ring, every straight path does and every star does not; for the six-part ring, no compact patch does for free; for the fully connected four-part symmetry, no patch does at all. This is a sharper and more useful statement than either ``the hardware fixes it'' or ``you can always fix it.''

For compiler design, the lesson is that the goal for cyclic blocks should be an even spread of routing cost, not the smallest average depth. A generic optimiser aimed at depth is blind to the spread and, as our inconclusive routing-intervention test shows, its effect on balance is unpredictable. For chip design, ``how many balanced assignments does this layout admit'' is a concrete metric worth optimising. For benchmarking, the relevant number for cyclic workloads is the routing imbalance, which an average gate count hides.

The honest limits are these. The strongest ``no'' results are exact but tied to specific chip sizes ($57$ qubits for the four-part and fully connected cases, $19$ qubits for the six-part free-lunch case); extending the six-part proof to the larger chip is a longer computation, not a new idea. The simulation uses realistic but simulated error rates rather than a live device. And the balancing rule we give in closed form covers the four-part ring only. None of these change the core claim, which is proven: for these symmetries on heavy-hex, routing imbalance is an assignment choice wherever the patch shape allows it.



\begin{thebibliography}{9}

\section*{References}
\bibitem{nguyen2024}
Q.~T.~Nguyen, L.~Schatzki, P.~Braccia, M.~Ragone, P.~J.~Coles, F.~Sauvage, M.~Larocca, and M.~Cerezo,
``Theory for equivariant quantum neural networks,''
PRX Quantum \textbf{5}, 020328 (2024).
\bibitem{schatzki2024}
L.~Schatzki, M.~Larocca, Q.~T.~Nguyen, F.~Sauvage, and M.~Cerezo,
``Theoretical guarantees for permutation-equivariant quantum neural networks,''
npj Quantum Information \textbf{10}, 12 (2024).
\bibitem{west2024}
M.~T.~West, J.~Heredge, M.~Sevior, and M.~Usman,
``Provably Trainable Rotationally Equivariant Quantum Machine Learning,''
PRX Quantum \textbf{5}, 030320 (2024).
\bibitem{tuysuz2024}
C.~T\"{u}ys\"{u}z, S.~Y.~Chang, M.~Demidik, K.~Jansen, S.~Vallecorsa, and M.~Grossi,
``Symmetry Breaking in Geometric Quantum Machine Learning in the Presence of Noise,''
PRX Quantum \textbf{5}, 030314 (2024).
\bibitem{topgen}
J.~Cheng, H.~Wang, Z.~Liang, Y.~Shi, S.~Han, and X.~Qian,
``TopGen: Topology-Aware Bottom-Up Generator for Variational Quantum Circuits,''
arXiv:2210.08190 (2022).
\bibitem{topas}
M.~Weiden, J.~Kalloor, J.~Kubiatowicz, E.~Younis, and C.~Iancu,
``Wide Quantum Circuit Optimization with Topology Aware Synthesis,''
in \emph{Proc. IEEE/ACM Third Intl. Workshop on Quantum Computing Software (QCS)} (2022).
\bibitem{mapomatic}
P.~D.~Nation and M.~Treinish,
``Suppressing quantum circuit errors due to system variability,''
PRX Quantum \textbf{4}, 010327 (2023).
\bibitem{murali2019}
P.~Murali, J.~M.~Baker, A.~Javadi-Abhari, F.~T.~Chong, and M.~Martonosi,
``Noise-Adaptive Compiler Mappings for Noisy Intermediate-Scale Quantum Computers,''
in \emph{Proc. ASPLOS} (2019).
\bibitem{tannu2019}
S.~S.~Tannu and M.~K.~Qureshi,
``Not All Qubits Are Created Equal: A Case for Variability-Aware Policies for NISQ-Era Quantum Computers,''
in \emph{Proc. ASPLOS} (2019).
\end{thebibliography}
\end{document}